\documentclass[aps,prc,showpacs,preprintnumbers,superscriptaddress,nofootinbib,floatfix,10pt]{revtex4-1}
\usepackage{amsfonts,amssymb,stmaryrd,latexsym,amsmath}
\usepackage{subfigure,graphicx}
\usepackage{times}
\usepackage{slashed}
\usepackage[colorlinks]{hyperref}

\usepackage{graphicx}
\usepackage{amsmath, amssymb, bm}
\usepackage{dcolumn}
\usepackage{bm}
\usepackage{comment}
\usepackage{tabularx}
\usepackage{multirow}

\begin{document}

\title{Galilean invariance restoration on the lattice}

\author{Ning~Li} 
\affiliation{Facility for Rare Isotope Beams and Department of Physics and
Astronomy, Michigan State
University, MI 48824, USA}

\author{Serdar~Elhatisari}
\affiliation{Helmholtz-Institut f\"ur Strahlen- und
             Kernphysik and Bethe Center for Theoretical Physics,
             Universit\"at Bonn,  D-53115 Bonn, Germany}
\affiliation{Faculty of Engineering, Karamanoglu Mehmetbey University, Karaman 70100, Turkey}

\author{Evgeny~Epelbaum}
\affiliation{Ruhr University Bochum, Faculty of Physics and Astronomy, Institute
for Theoretical Physics II, 
D-44870 Bochum, Germany}

\author{Dean~Lee}
\affiliation{Facility for Rare Isotope Beams and Department of Physics and Astronomy, Michigan State
University, MI 48824, USA}
\affiliation{Department~of~Physics, North~Carolina~State~University, Raleigh,
NC~27695, USA}
                     
\author{Bingnan~Lu}
\affiliation{Facility for Rare Isotope Beams and Department of Physics and
Astronomy, Michigan State
University, MI 48824, USA}

\author{Ulf-G.~Mei{\ss}ner}  
\affiliation{Helmholtz-Institut f\"ur Strahlen- und
             Kernphysik and Bethe Center for Theoretical Physics,
             Universit\"at Bonn,  D-53115 Bonn, Germany}  
\affiliation{Institute~for~Advanced~Simulation, Institut~f{\"u}r~Kernphysik,
and
J{\"u}lich~Center~for~Hadron~Physics,~Forschungszentrum~J{\"u}lich,
D-52425~J{\"u}lich, Germany}     
\affiliation{Tbilisi State University, 0186 Tbilisi, Georgia}

\begin{abstract}
We consider the breaking of Galilean invariance due to  different lattice cutoff effects
in moving frames and  a nonlocal smearing parameter which is used in the construction of the nuclear lattice interaction.  
The dispersion relation and neutron-proton scattering phase shifts are used  to investigate
the Galilean invariance breaking effects and ways to restore it. 
For $S$-wave channels, ${}^1S_0$ and ${}^3S_1$, we present the neutron-proton scattering phase shifts in moving frames 
calculated using both L{\"u}scher's formula and the spherical wall method, as well as the dispersion relation. 
For the $P$ and $D$ waves, we present the neutron-proton scattering phase shifts in moving frames  calculated using the 
spherical wall method. We find that the Galilean invariance breaking effects stemming from the lattice artifacts partially 
cancel those caused by the nonlocal smearing parameter. Due to this cancellation, the Galilean invariance breaking 
effect is small, and the Galilean invariance can be restored by introducing  Galilean invariance restoration operators. 

\end{abstract}


\maketitle

\section{INTRODUCTION}
Chiral effective field theory  (EFT) allows one to construct the nuclear force systematically in powers
of $Q/\Lambda_{\chi}$, where $Q$ is a soft scale (pion mass, transferred momenta, etc), while  
$\Lambda_{\chi}$ ($\approx$ 0.6~GeV) 
is the pertinent hard scale~\cite{Weinberg:1991um,Epelbaum:1998ka,Epelbaum:1999dj,Epelbaum:2004fk}.
In chiral EFT, the most important contribution appears at leading order (LO) or order $(Q/\Lambda_\chi)^0$, the second most 
important contribution at next-to-leading order (NLO) or order $(Q/\Lambda_\chi)^2$, the third most important 
contribution at next-to-next-to-leading order (N2LO) or order $(Q/\Lambda_\chi)^3$, and so on. According to the 
power counting of 
chiral EFT, the LO nucleon-nucleon (NN) interaction includes the static one-pion-exchange potential as well as 
momentum independent contact interactions, the NLO 
NN interaction includes the leading two-pion-exchange potential (TPEP) and contact interactions with two derivatives,  
the N2LO interaction includes only  corrections to the 
TPEP, and the N3LO NN interaction includes further corrections to the OPEP and sub-leading TPEP as well as contact interactions
with four derivatives. See~\cite{Epelbaum:2008ga,Machleidt:2011zz} for review 
papers on chiral nuclear EFT. 
  
In the past decades, nuclear lattice effective field theory  (NLEFT) combining Monte Carlo simulations 
on a space-time grid and nuclear forces derived within chiral EFT 
has become a powerful tool for {\it ab initio} calculations of the few- and many-body problems. NLEFT has been widely used to 
study nuclear structure~\cite{Epelbaum:2011md,Epelbaum:2012qn,Epelbaum:2013paa}  and nuclear reactions~\cite{Elhatisari:2015iga}. 
See~\cite{Lee:2008fa} for an early review article.  Since NLEFT is powerful for {\it ab initio} calculations, getting 
an efficient and precise nuclear force is particularly important,  
which is a more difficult task than in the continuum due to the lattice artifacts stemming from the nonzero lattice spacing. 
To reduce the lattice artifacts, non-locally smeared operators were introdued in~\cite{Elhatisari:2016owd}. With only a few parameters, 
the binding energies of nuclei with nucleons $A \leq 20$ are produced with good precision. In Ref.~\cite{Li:2018ymw} ,  these non-locally smeared 
operators were extended up to next-to-next-to-next-to-leading order (N3LO) in chiral EFT for neutron-proton scattering.  

However, in a lattice-regularized system, finite-lattice spacing effects are 
different in moving frames. This breaks the Galilean invariance~\cite{Lee:2007jd}, 
which is the statement that the laws of Newtonian physics for a non-relativistic system are independent of the velocity of the center of mass. 
There is also some breaking of Galilean invariance caused by the nonlocal smearing parameter $s_{\rm NL}$ we use in the construction of the lattice 
interaction as it induces the explicit 
dependence of the lattice interaction on the momentum of the center of mass. 
In the present work, we focus on the lattice calculations with lattice
spacing $a = 1.32$~fm and the N3LO nucleon-nucleon interactions from~\cite{Li:2018ymw}. 
We quantify the effects of Galilean invariance breaking by analyzing the dispersion relation and  neutron-proton scattering 
phase shifts in moving frames.
We also show how to restore the Galilean invariance 
by including the contribution of the Galilean invariance restoration operators.  This is the main finding of this paper that will
be used in future NLEFT investigations.

 The paper is organized as follows. After the introduction, in section~\ref{formalism}  we will present the formalism including the 
 lattice nucleon-nucleon interaction up to N3LO in chiral EFT, the L{\"u}scher's formula and spherical wall method used to 
 extract the scattering phase shifts. Then, we present the numerical results and make discussions in section~\ref{numericalResults}. 
 Finally, we summarize our results in section~\ref{summary}.

\section{FORMALISM}\label{formalism}

Before present the details of our formalism, it is useful to state some conventions appearing many times in the present paper. 
Throughout this work we use $a$ for the spatial lattice spacing, $L$  denotes the number of lattice points in each spacial direction, 
and ${\bf P}$ is the momentum of the center of mass. All parameters and operators are first expressed in lattice units, 
which correspond to the physical values multiplied by appropriate powers of $a$. Our final results are presented in physical units.

Different from our previous calculations, where the transfer matrix formalism was used, here we utilize the Hamiltonian formalism since the 
transfer matrix formalism can induce additional breaking of Galilean invariance due to the nonzero temporal lattice spacing. In our calculation, 
the Hamiltonian has the form, 
\begin{eqnarray}
H = H_{\rm free} + V_{\rm 2N}^{\rm short} + V_{\rm 2N}^{\rm long}.
\end{eqnarray}
For the free Hamiltonian we use an $O(a^4)$-improved action of the form  \cite{Lee:2008fa},
\begin{align}
H_{\rm free}= &\frac{49}{12m_N}\sum_{\bf n} a^\dagger({\bf n}) a({\bf n})-\frac{3}{4m_N}\sum_{{\bf n},i}
\sum_{\langle{\bf n'}\,{\bf n}\rangle_i} a^\dagger({\bf n'}) a({\bf n}) \nonumber
\\
&+\frac{3}{40m_N}\sum_{{\bf n},i}\sum_{\langle\langle{\bf n'}\,{\bf n}\rangle\rangle_i} a^\dagger({\bf n'})
a({\bf n})-\frac{1}{180m_N}\sum_{{\bf n},i}
\sum_{\langle\langle\langle{\bf n'}\,{\bf n}\rangle\rangle\rangle_i} a^\dagger({\bf n'}) a({\bf n}),
\end{align}
where $a^\dag$ and $a$ are the creation and annihilation operators for a nucleon, respectively, and $m_N$ denotes the nucleon mass.
The number of brackets under the sum refer to the nearest, next-to-nearest and next-to-next-to-nearest neighbors used in the
representation of the derivatives.
$V_{\rm 2N}^{\rm short}$  is the short-range nucleon-nucleon interaction accounted by contact interactions while $V_{\rm 2N}^{\rm long}$ 
denotes the long-range NN interaction provided by the pion-exchange potentials.  

\subsection{Nucleon-nucleon interaction on the lattice }
Up to N3LO in chiral EFT,  the short-range nucleon-nucleon interaction includes contact interactions at LO, NLO and N3LO, 
\begin{eqnarray}
V_{\rm 2N}^{\rm short} = V_{\rm contact}^{(Q/\Lambda_\chi)^0} + V_{\rm contact}^{(Q/\Lambda_\chi)^2} + V_{\rm contact}^{(Q/\Lambda_\chi)^4}~. 
\end{eqnarray}
At LO, two non-locally smeared contact operators were introduced in Ref.~\cite{Li:2018ymw}. These read 
\begin{eqnarray}
V_{{}^1S_0, (Q/\Lambda_\chi)^0}  &=& \sum_{I_z=-1,0,1}\left[ O^{0,s_{NL}}_{0,0,0,0,1,I_z}({\bf n})\right]^{\dagger}O^{0,s_{NL}}_{0,0,0,0,1,I_z}({\bf n}),
\end{eqnarray}
for the ${}^1S_0$ channel, and 
\begin{eqnarray}
V_{{}^3S_1, (Q/\Lambda_\chi)^0} &=& \sum_{J_z=-1,0,1}\left[ O^{0,s_{NL}}_{1,0,1,J_z,0,0}({\bf n})\right]^{\dagger}O^{0,s_{NL}}_{1,0,1,J_z,0,0}({\bf n}),
\end{eqnarray}
for the ${}^3S_1$ channel. We refer to App.~\ref{LatticeDefinition} for the definitions of the pair creation operator $O^\dag$ and 
pair annihilation operator $O$. The contact operators at NLO  and N3LO can be written in a similar manner. 
Their specific expressions which are not given here for simplicity can be found in~\cite{Li:2018ymw}.

Additionally, we also include an SU(4)-invariant short-range operator at LO, which has been shown 
to be important for nuclear binding~\cite{Elhatisari:2016owd,Lu:2018bat}, 
\begin{equation}
V_0 = \frac{C_0}{2}: \! \! \! \sum_{{\bf n'},{\bf n},{\bf n''}}\sum_{i',j'} a^{s_{\rm NL}\dagger}_{i',j'}({\bf n'})a^{s_{\rm NL}}_{i',j'}({\bf n'})
f_{s_{\rm L}}({\bf n'} - {\bf n})f_{s_{\rm L}}({\bf n} - {\bf n''})
\sum_{i'',j''} a^{s_{\rm NL}\dagger}_{i'',j''}({\bf n''})a^{s_{\rm NL}}_{i'',j''}({\bf n''}):,
\end{equation}
where $::$ denotes normal ordering, and the local smearing function $f_{s_{\rm L}}({\bf n})$ is defined as
\begin{eqnarray}
f_{\rm s_L} = 
\left\{ 
\begin{array}{cc}
1, & |{\bf n}| = 0, \\
s_L , & |{\bf n}| = 1, \\
0, & {\rm otherwise}. \\
\end{array}
\right.
\end{eqnarray}
The index $i$ corresponds to nucleon spin, and the index $j$ corresponds to nucleon isospin. 
The dressed creation operator $a^{s_{\rm NL} \dag}$ and annihilation operator $a^{s_{\rm NL}}$ are defined respectively as 
 \begin{equation}
 a^{s_{\rm NL}}_{i,j}({\bf n})=a_{i,j}({\bf n})+s_{\rm NL}\sum_{|{\bf n'}|=1}a_{i,j}({\bf n}+{\bf n'})~,
\end{equation}
and 
\begin{equation}
a^{s_{\rm NL}\dagger}_{i,j}({\bf n})=a^{\dagger}_{i,j}({\bf n})+s_{\rm NL}\sum_{|{\bf n'}|=1}a^{\dagger}_{i,j}({\bf n}+{\bf n'})~.
\end{equation}
We use the dressed creation (annihilation) operator to create (annihilate)  the nucleon placed at the exact lattice site 
as well as the nucleon located at its nearest-neighbor lattice sites. In this manner, some of the lattice artifacts induced by the nonzero lattice 
spacing can be removed. However, the nonzero value of $s_{\rm NL}$ leads to a breaking of Galilean invariance because 
it makes the NN interaction depend on the velocity of the center of mass.

For the long-range interaction, we include the one-pion-exchange potential (OPEP) at LO, and the  two-pion-exchange 
potentials (TPEP) at NLO, N2LO, and N3LO. 
\begin{eqnarray}
V_{\rm 2N}^{\rm long} = V_{\rm OPE}^{(Q/\Lambda_\chi)^0} + V_{\rm TPE}^{(Q/\Lambda_\chi)^2} + V_{\rm TPE}^{(Q/\Lambda_\chi)^3} 
+ V_{\rm TPE}^{(Q/\Lambda_\chi)^4}.
\end{eqnarray}
The one-pion exchange potential  $V_{\rm OPE}$ has the form
\begin{align}
V_{\rm OPE}=-\frac{g_A^2}{8F^2_{\pi}}\sum_{{\bf n',n},S',S,I}
:\rho_{S',I\rm }({\bf n'})f_{S'S}({\bf n'}-{\bf n})\rho_{S,I}({\bf n}):,
\end{align}
where $f_{S'S}$ is defined as
\begin{align}
f_{S'S}({\bf n'}{\bf -n})=\frac{1}{L^3}\sum_{\bf q}\frac{Q(q_{S'})Q(q_{S})\exp[-i{\bf q}\cdot({\bf
n'}-{\bf n})-b_{\pi}({\bf q}^2+M^2_{\pi})]}{{\bf q}^2 + M_{\pi}^2},\label{OPE}
\end{align}
and each lattice momentum component $q_S$ is an integer multiplied by $2\pi/L$.
The function $Q(q_S)$ is given by
\begin{equation}
Q(q_S) = \frac{3}{2}\sin(q_S)-\frac{3}{10}\sin(2q_S)+\frac{1}{30}\sin(3q_S), \label{momentum:q}
\end{equation}
which equals $q_S$ up to correction of order $q_S^7$. We use the definition of Eq.~(\ref{momentum:q}) for the 
nucleon momentum on the lattice to remove the finite lattice volume effects. We include the parameter $b_\pi$ to regularize the short-range behavior 
of the one-pion-exchange potential. As in previous calculations, we set $b_\pi = 0.25$ in lattice units. For calculations with 
lattice spacing $a = 1.32$ fm, this is equivalent to $\Lambda = 300$ MeV in the form factor 
\begin{eqnarray}
F({\bf q}) = \exp\left[-\frac{({\bf q}^2 + m_{\pi}^2)}{\Lambda^2}\right].
\end{eqnarray} 
We use the combination ${\bf q}^2+M^2_{\pi}$ in the exponential as suggested in \cite{Reinert:2017usi} 
as a momentum-space regulator
which does not affect the long-distance behavior of the pion-exchange potential. 

As we are solving non-relativistic Schr{\"o}dinger equations, we neglect the relativistic corrections to the NN force at N3LO stemming 
from the $1/m_N^2$-corrections to the static OPEP and  $1/m_N$-corrections to static TPEP including spin-orbital interacting 
terms~\cite{Epelbaum:2014efa,Reinert:2017usi}. 
As a result, the long-range pion-exchange potential is totally local and independent of the velocity of the center of mass. 
Therefore, this part  does not break Galilean invariance. Instead of providing the lengthy expressions of TPEP, we refer  the reader
to~\cite{Kaiser:2001pc,Epelbaum:2014efa,Entem:2014msa,Reinert:2017usi} for the specific expressions. 
 
\subsection{Galilean invariance restoration operators}
To restore the Galilean invariance for the two-nucleon system, we introduce the pair hopping terms. 
We first illustrate with pointlike operators corresponding to the product  of total nucleon densities, 
 \begin{equation}
V_{\rm GIR} = V^0_{\rm GIR} + V^1_{\rm GIR} + V^2_{\rm GIR}, 
\end{equation}
where
\begin{align}
V^0_{\rm GIR} =C_{\rm GIR}^0 \sum_{{\bf n},i,j,i',j'} a^{\dagger}_{i,j}({\bf n})a^{\dagger}_{i',j'}({\bf n})a_{i',j'}({\bf n})a_{i,j}({\bf n})\label{GIR0}
\end{align}
denotes no hopping, 
\begin{align}
V^1_{\rm GIR} = C_{\rm GIR}^1\sum_{{\bf n},i,j,i',j'}\sum_{\bf |n'| = 1}  a^{\dagger}_{i,j}({\bf n}+{\bf n'})
a^{\dagger}_{i',j'}({\bf n}+{\bf n'}) a_{i',j'}({\bf n})a_{i,j}({\bf n})\label{GIR1}
\end{align}
is the nearest-neighbor hopping term, and 
\begin{align}
V^2_{\rm GIR} = C_{\rm GIR}^2\sum_{{\bf n},i,j,i',j'}\sum_{\bf |n'| = \sqrt{2}}  a^{\dagger}_{i,j}({\bf n}+{\bf n'})
a^{\dagger}_{i',j'}({\bf n}+{\bf n'})a_{i',j'}({\bf n})a_{i,j}({\bf n}) \label{GIR2}
\end{align}
is the next-to-nearest-neighbor hopping term for the nucleon-nucleon pair. 

Let us write $\lvert {\bf P} \rangle$ as a two-body bound-state wave function with total momentum ${\bf P}$.  
We note that $\langle{{\bf P}} \vert V^0_{\rm GIR} \vert {\bf P} \rangle\\ $ is independent of ${\bf P}$, and so we have 
\begin{equation}
\langle{{\bf P}} \vert V^0_{\rm GIR} \vert {\bf P} \rangle =C_{\rm GIR}^0\langle {\bf 0} \vert V^0_{\rm GIR} \vert {\bf 0} \rangle,
\end{equation}
where $\lvert {\bf 0}\rangle$ is the two-body bound-state wave function with zero total momentum.  Furthermore,
\begin{align}
\langle{ {\bf P}} & \vert V^1_{\rm GIR} \vert {\bf P} \rangle = 2C_{\rm GIR}^1\left[\cos(P_x) + \cos(P_y) + \cos(P_z) \right]
\langle {\bf 0} \vert V^0_{\rm GIR} \vert {\bf 0} \rangle~, 
\end{align}
and 
\begin{align}
\langle{ {\bf P}} & \vert V^2_{\rm GIR} \vert {\bf P} \rangle = 4\big[\cos(P_x) \cos(P_y) + \cos(P_y)\cos(P_z)
 + \cos(P_z)\cos(P_x)\big] \langle {\bf 0} \vert V^0_{\rm GIR} \vert {\bf 0} \rangle~.
\end{align}
Combining the hopping term with the contact terms we can construct the GIR operators. For simplicity, we only 
take the lowest-order contact operator of each channel to construct the GIR operators. For example, 
the GIR operator for the ${}^1S_0$ channel reads
\begin{align}
V^{{}^1S_0}_{\rm GIR} =~& C_{{\rm GIR}, 0}^{{}^1S_0} \sum_{\bf n}\sum_{I_z = -1, 0, 1}\left[O_{0, 0, 0, 0, 1, I_z}^{0, s_{NL}}({\bf n})\right]^\dag 
O_{0, 0, 0, 0, 1, I_z}^{0, s_{NL}}({\bf n}) \nonumber \\
& +C_{{\rm GIR}, 1}^{{}^1S_0} \sum_{\bf n}\sum_{|{\bf n}^\prime| = 1} \sum_{I_z = -1, 0, 1}
\left[O_{0, 0, 0, 0, 1, I_z }^{0, s_{NL}}({\bf n} + {\bf n}^\prime)\right]^\dag
O_{0, 0, 0, 0, 1, I_z}^{0, s_{NL}}({\bf n}) \nonumber \\
& + C_{{\rm GIR}, 2}^{{}^1S_0} \sum_{\bf n}\sum_{|{\bf n}^\prime|  = \sqrt{2} } \sum_{I_z = -1, 0,1}
\left[O_{0, 0, 0, 0, 1, I_z}^{0, s_{NL}}({\bf n} + {\bf n}^\prime)\right]^\dag
O_{0, 0, 0, 0, 1, I_z}^{0, s_{NL}}({\bf n}),
\end{align} 
whereas that for the ${}^1P_1$ channel is 
\begin{align}
V^{{}^1P_1}_{\rm GIR} =~&C_{{\rm GIR}, 0}^{^{1P_1}}\sum_{\bf n} \sum_{J_z= -1, 0,1}
\left[O_{0, 1, 1, J_z, 0, 0}^{0, s_{NL}}({\bf n})\right]^\dag O_{0, 1, 1, J_z, 0, 0}^{0, s_{NL}}({\bf n}) \nonumber \\
& + C_{{\rm GIR}, 1}^{{}^1P_1} \sum_{\bf n} \sum_{|{\bf n}^\prime| = 1} \sum_{J_z = -1, 0, 1} 
\left[O_{0, 1, 1, J_z, 0, 0}^{0, s_{NL}}({\bf n} + {\bf n}^\prime)\right]^\dag 
O_{0, 1, 1, J_z, 0, 0}^{0, s_{NL}}({\bf n}) \nonumber \\
& + C_{{\rm GIR}, 2}^{{}^1P_1} \sum_{\bf n} \sum_{|{\bf n}^\prime| = \sqrt{2}} \sum_{J_z = -1, 0, 1} 
\left[O_{0, 1, 1, J_z, 0, 0}^{0, s_{NL}}({\bf n} + {\bf n}^\prime)\right]^\dag
O_{0, 1, 1, J_z, 0, 0}^{0, s_{NL}}({\bf n}).
\end{align}
Using these GIR operators, we can restore Galilean invariance for each channel by finely tuning 
$C_{{\rm GIR}, i} (i = 0, 1, 2)$ with the constraint, 
\begin{eqnarray}
C_{{\rm GIR}, 0} + 6C_{{\rm GIR}, 1} + 12 C_{{\rm GIR}, 2} = 0, 
\end{eqnarray} 
which is the requirement that the GIR correction should be vanishing for zero total momentum. 
Specifically, we take the Nijmegen phase shifts as input to determine the LECs for each channel in the rest frame, 
and then determine the coefficients $C_{{\rm GIR}, i}$ by fitting the phase shifts 
in the boosted frames, where the lattice results in the rest frame are taken as input. 
For example, two LECs for ${}^1P_1$ are fixed at N3LO without GIR, then two additional coefficients, $C_{{\rm GIR}, i}$, are used to restore the Galilean invariance.

\subsection{L{\"u}scher's formula}

In \cite{Luscher:1990ux}, L{\"u}scher derived a simple formula connecting the two-body S-wave scattering phase 
shift $\delta_0$ with the energy levels calculated  
in the lattice framework. It reads  
\begin{eqnarray}
\exp\left(2i\delta_0(k)\right) = \frac{\zeta_{00}(1;q^2) + i \pi^{3/2}q}{\zeta_{00}(1;q^2)-i\pi^{3/2}q}, \label{static}
\end{eqnarray}
where 
\begin{eqnarray}
q = \frac{kL}{2\pi}, 
\end{eqnarray}
and 
\begin{eqnarray}
\zeta_{00}(s;q^2) = \frac{1}{\sqrt{4\pi}}\sum_{{\bf n} \in  Z^3} ({\bf n}^2 - q^2)^{-s}
\end{eqnarray}
is the zeta function which is convergent when ${\rm Re}(s) >3/2$, and can be analytically continued to  
$s = 1$. Then, this formula was  generalized to moving frames with center-of-mass momentum 
${\bf P} = (2\pi/L) {\bf k}$~\cite{Rummukainen:1995vs,Kim:2005gf,Beane:2003da,Feng:2004ua}, 
\begin{eqnarray}
\delta_0(k) = \arctan\left(\frac{\gamma q \pi^{3/2}} {\zeta_{00}^{\bf d}(1;q^2)}\right), \label{moving}
\end{eqnarray}
where 
\begin{eqnarray}
\zeta_{00}^{\bf d}(s;q^2) = \frac{1}{\sqrt{4\pi}}\sum_{{\bf r}\in P_d}({\bf r}^2 - q^2)^{-s},
\end{eqnarray}
is the generalized zeta function. The summation region $P_{\bf d}$ is defined as 
\begin{eqnarray}
P_{\bf d} = \left\{{\bf r} \in R^3|{\bf r} = \gamma^{-1}({\bf n } + {\bf d}/2), {\bf n } \in Z^3 \right\}, 
\end{eqnarray}
where $\gamma$  is the Lorentz factor and $\gamma^{-1} {\bf n}$ is the
shorthand notation for 
$ \gamma^{-1} {\bf n}_{\parallel} + {\bf n}_{\perp}$. It is easy to check that formulae Eq.~(\ref{static}) and~(\ref{moving}) are the same when ${\bf P} = 0$. 
The expressions for the numerical calculation of the generalized zeta function can be found in Refs.~\cite{Rummukainen:1995vs,Gockeler:2012yj}.
Refer to \cite{Briceno:2012yi,Briceno:2013bda,Briceno:2013lba,Gockeler:2012yj} for several interesting lattice QCD calculations in the moving frames.  

In our calculation, the L{\"u}scher formula is applied to calculate the neutron-proton scattering phase shifts for only  the S-wave channels. 
This is done  because L{\"u}scher's formula is not an efficient method to extract the scattering phase shifts for the $P$, $D$ and higher partial waves.
Even for ${}^3S_1$, we find a small discrepancy between the results using L{\"u}scher's formula and those using the spherical wall method. This is because there 
is a systematic error in the mixing of different channels when using the L{\"u}scher's formula. We will come back to this later. 

\subsection{Spherical wall method} 
In addition to  L{\"u}scher's formula, the spherical wall method is another approach to extract the scattering phase shifts. 
Differently from L{\"u}scher's formula connecting the scattering phase shifts with the energy levels, the spherical wall method extracts the 
scattering phase shifts from the wave function. To calculate the scattering phase shifts and mixing angles using the spherical 
wall method, we first construct radial wave functions in moving frame with momentum ${\bf P}$ through the spherical harmonics with quantum numbers $(l, l_z)$ \cite{Lu:2015riz, Elhatisari:2016hby}, 
\begin{eqnarray}
\left|r\right>^{l, l_z}_{\bf P} = \sum_{\hat r^\prime}  \exp(-{i\bf P}\cdot {\bf r}^\prime )Y_{l, l_z}({\hat r}^\prime) \delta_{|{\bf r}^\prime| = r} \left|{\bf r}^\prime\right>,
\end{eqnarray}
where ${\hat r}^\prime$ runs over all lattice sites  having the same radial lattice distance, and ${\bf P} = (2\pi/L) {\bf k}$ is the quantized 
center-of-mass momentum on the lattice. Using this definition for the radial wave function, the Hamiltonian matrix  over a three-dimensional 
lattice can be reduced to a one-dimensional radial Hamiltonian, $H_{{\bf r}, {\bf r}^\prime} \rightarrow H_{r, r^\prime}$.

After solving the Sch{\"o}dinger equation,  the phase shifts and mixing angles can be extracted from the radial wave function in the region where 
the NN force is vanishing. In this range, the wave function is a
superposition of the incoming plane wave and outgoing radial wave which can be expanded as~\cite{Lu:2015riz,Li:2018ymw}
\begin{eqnarray}
\langle r | k, l\rangle =  A_j h_l^{(1)}(kr) + B_j h_l^{(2)}(kr), 
\end{eqnarray}
where $h_l^{(1)}(kr)$ and $h_l^{(2)}(kr)$ are the spherical Hankel functions.
 $k = \sqrt{2\mu E}$ with $\mu$ the reduced mass and $E$ the relative energy of the two-nucleon system. 
 The scattering coefficients $A_j$ and $B_j$ satisfy the 
relations, 
\begin{eqnarray}
B_j = S_jA_j,  \label{eq:ph}
\end{eqnarray}
where $S_j = \exp\left(2i \delta_j\right)$ is the $S$-matrix and $\delta_j$ is the phase shift. The phase shift  is determined by setting 
\begin{eqnarray}
\delta_j = \frac{1}{2i}\log\left(\frac{B_j}{A_j}\right)~.
\end{eqnarray}
In the case of the coupled channels with $j > 0$, both of the coupled partial waves, $l = j -1$ and $l = j + 1$, satisfy 
Eq.~(\ref{eq:ph}), and the $S$-matrix couples the two 
channels together. Throughout this work  we adopt the so-called  Stapp parameterization of the phase shifts and mixing angles for the coupled channels \cite{Stapp:1956mz}, 
\begin{eqnarray}
S = \left[
\begin{array}{cc}
\cos(2\epsilon) \exp \left(2i\delta_{j-1}^{1j} \right)   &    i\sin(2\epsilon) \exp \left(i \delta_{j-1}^{1j} + i \delta_{j+1}^{1j} \right) \\
i \sin{2\epsilon} \exp \left(i\delta_{j-1}^{1j} + i \delta_{j+1}^{1j} \right)   &  \cos(2\epsilon) \exp \left( 2i \delta_{j+1}^{1j} \right) \\
\end{array}
\right].
\end{eqnarray}

\section{NUMERICAL RESULTS AND DISCUSSION} \label{numericalResults}

In our calculation, we first determine the low-energy constants by matching the calculated neutron-proton 
scattering phase shifts to those from the Nijmegen partial analysis. Then, we boost the two-nucleon system to 
a moving frame with momenta ${\bf P} = (2\pi/L){\bf k}$ and calculate the phase shifts again. From the difference 
between these two results, we can read off the amount of the Galilean invariance breaking (GIB). We finally restore the 
Galilean invariance by tuning the coefficient $C_{{\rm GIR},i}$ to make the results independent of ${\bf P}$.
Since the dispersion relation is another good physical quantity to test the GIB, we also calculate it 
for both S-wave channels, ${}^1S_0$ and ${}^3S_1$.

For lattice parameters, we use the same values as those in one of our previous calculations in Ref.~\cite{Li:2018ymw}, namely, 
the spatial lattice spacing $a = 1.32$~fm, coefficient for the SU(4) contact potential $C_0 = -0.04455$~l.u. (lattice units),
local smearing parameter $s_L = 0.16985$~l.u., and nonlocal smearing parameter $s_{\rm NL} = 0.18566$~l.u..
We use $m_p = 938.272$~MeV and $m_n = 939.565$~MeV for the proton and the neutron mass, respectively. For the charged 
pion mass, we take $M_{\pi^{\pm}}$ = 139.57~MeV while for the neutral pion mass, we take $M_{\pi^0} = 134.97$~MeV. 
For the averaged pion mass we use $M_\pi = 138.03$~MeV.  Additionally, we use $F_{\pi} = 92.1$~MeV for the pion decay 
and $g_A = 1.287$ from the Goldberger-Treiman relation using the pion-nucleon coupling constant from Ref.~\cite{Baru:2011bw} 
for the nucleon axial coupling constant, respectively, and 
$c_1 = -1.10(3)$~GeV$^{-1}$, $c_2 = 3.57(4)$~GeV$^{-1}$, $c_3 = - 5.54(6)$~GeV$^{-1}$, and $c_4 = 4.17(4)$~GeV$^{-1}$~\cite{Hoferichter:2015tha},
for the low-energy constants appearing in the TPEP potentials. For the pion-nucleon LECs $d_i$ entering the chiral N3LO TPEP, we 
adopt $ \bar{d}_1 + \bar{d}_2 = 1.04$ GeV$^{-2}$, $\bar{d}_3 = -0.48$ GeV$^{-2}$, $\bar{d}_5= 0.14$ GeV$^{-2}$ 
and $\bar{d}_{14} - \bar{d}_{15} = -1.90$ GeV$^{-2}$ \cite{Reinert:2017usi}.

\subsection{Dispersion relation for the S waves}
\begin{figure}
\centering
\includegraphics[width=0.6\textwidth]{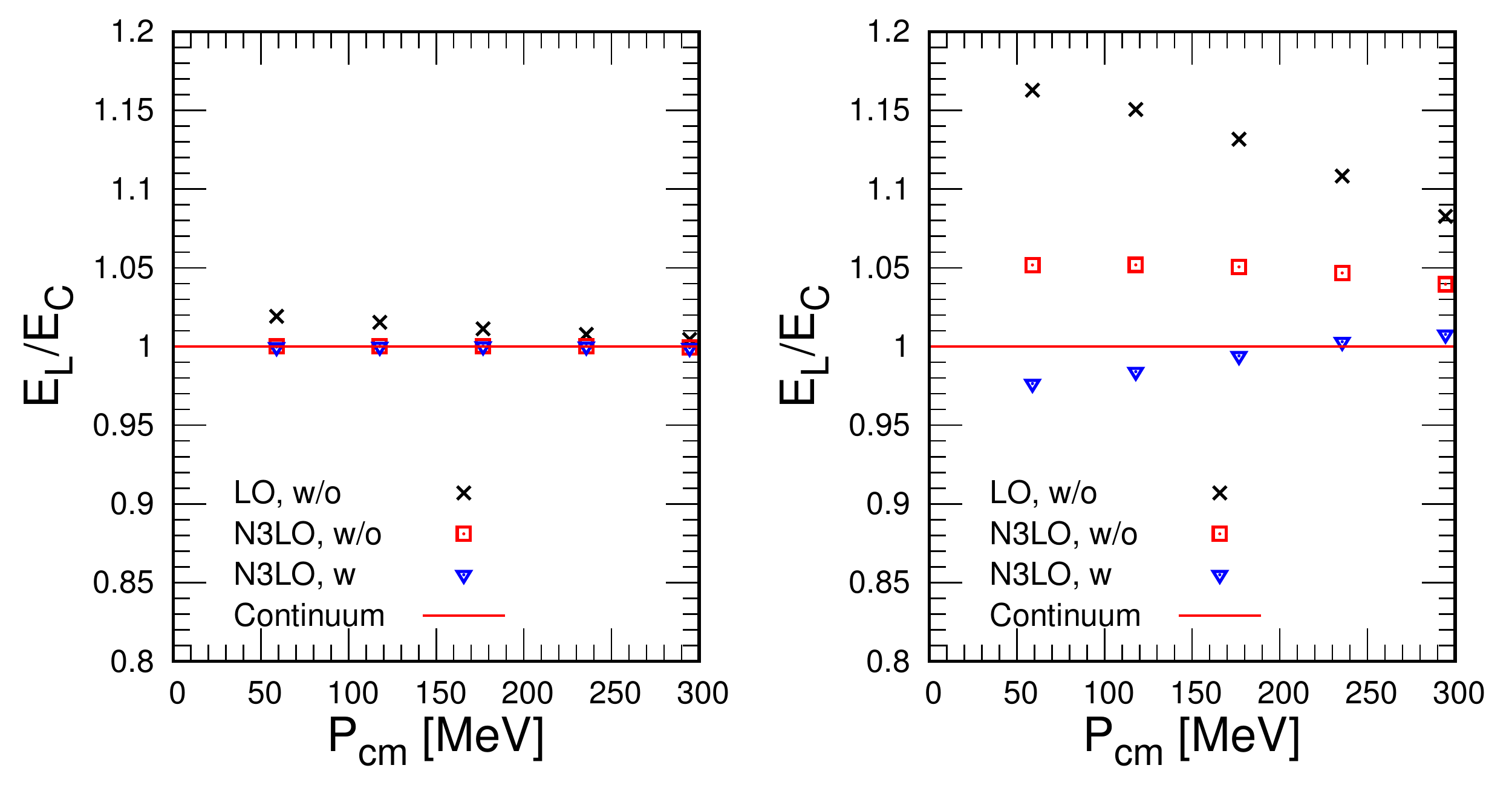}
\caption{(Color online) The ratios of the lattice and continuum energy as a function of the momentum of the center of mass.  
Left panel: ${}^1S_0$, right panel: ${}^3S_1$.  ``w/" means with GIR whereas ``w/o" means without GIR.}
\label{dp}
\end{figure}

We calculate the dispersion relation for the two $S$-wave channels,  ${}^1S_0$ and ${}^3S_1$, of the proton-neutron system in a cubic 
box of  volume $V = (32a)^3$ with  lattice spacing $a = 1.32$ fm. To make the effects better visible, we plot the ratios of the lattice and continuum 
energy as a function of the center-of-mass momentum. The results are shown in 
Fig.~\ref{dp}. $E_L/E_C$ is the ratio of the lattice and continuum energy.  
The left plot is for ${}^1S_0$ while the right plot gives ${}^3S_1$. We present the results without GIR at both LO and N3LO, 
which are used to read off the amount of Galilean invariance 
breaking. We also provide the results including GIR corrections at N3LO. 

From the plots, the lattice results for ${}^1S_0$ are closer to the continuum results that those for ${}^3S_1$. This is because the state we 
are boosting in the ${}^1S_0$ channel is a continuum state rather than bound state.
The almost perfect dispersion
suggests that it is not an efficient tool to investigating the GIB effect for ${}^1S_0$. 
Later, we will apply the proton-neutron scattering phase to study GIB in the ${}^1S_0$ channel. 
Differently from the ${}^1S_0$ case, the dispersion relation is very
useful to detect GIB in the ${}^3S_1$ channel as the ground state in
this case is a bound state. From the plots, it is  clear that compared to the LO result, the N3LO
values are closer to the continuum result. This indicates that there is less GIB effect for the N3LO interaction than for the LO interaction. 
Further, this indicates that GIB effect  stems from the nonlocal smearing parameter partially cancel those caused 
by the lattice artifacts since there are some  non-locally smeared contact terms at NLO and N3LO. 

\subsection{S-wave neutron-proton scattering phase shifts}

\begin{figure}
\centering
\begin{tabular}{cc}
\includegraphics[width=0.45\textwidth]{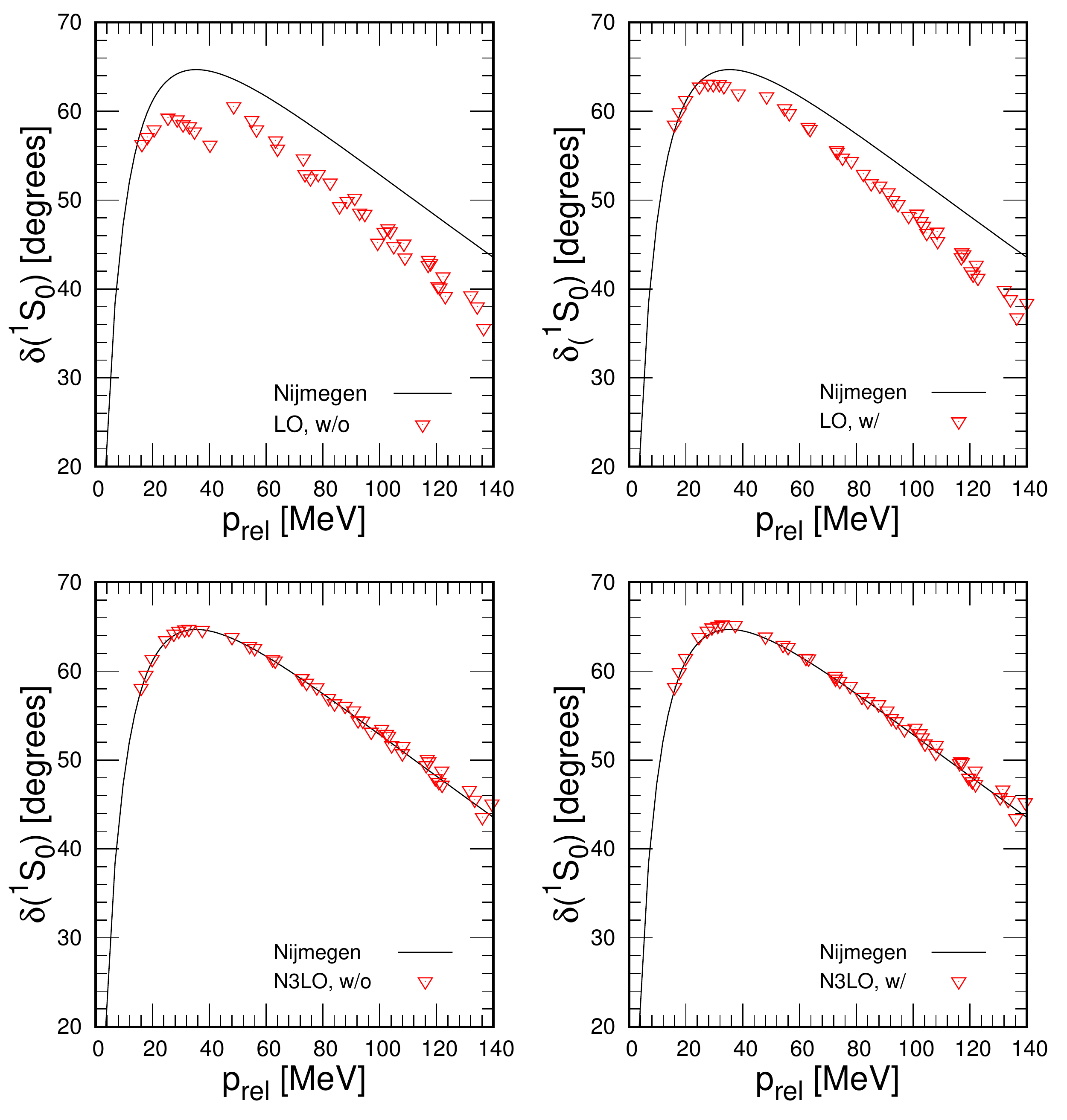}  &
\includegraphics[width=0.45\textwidth]{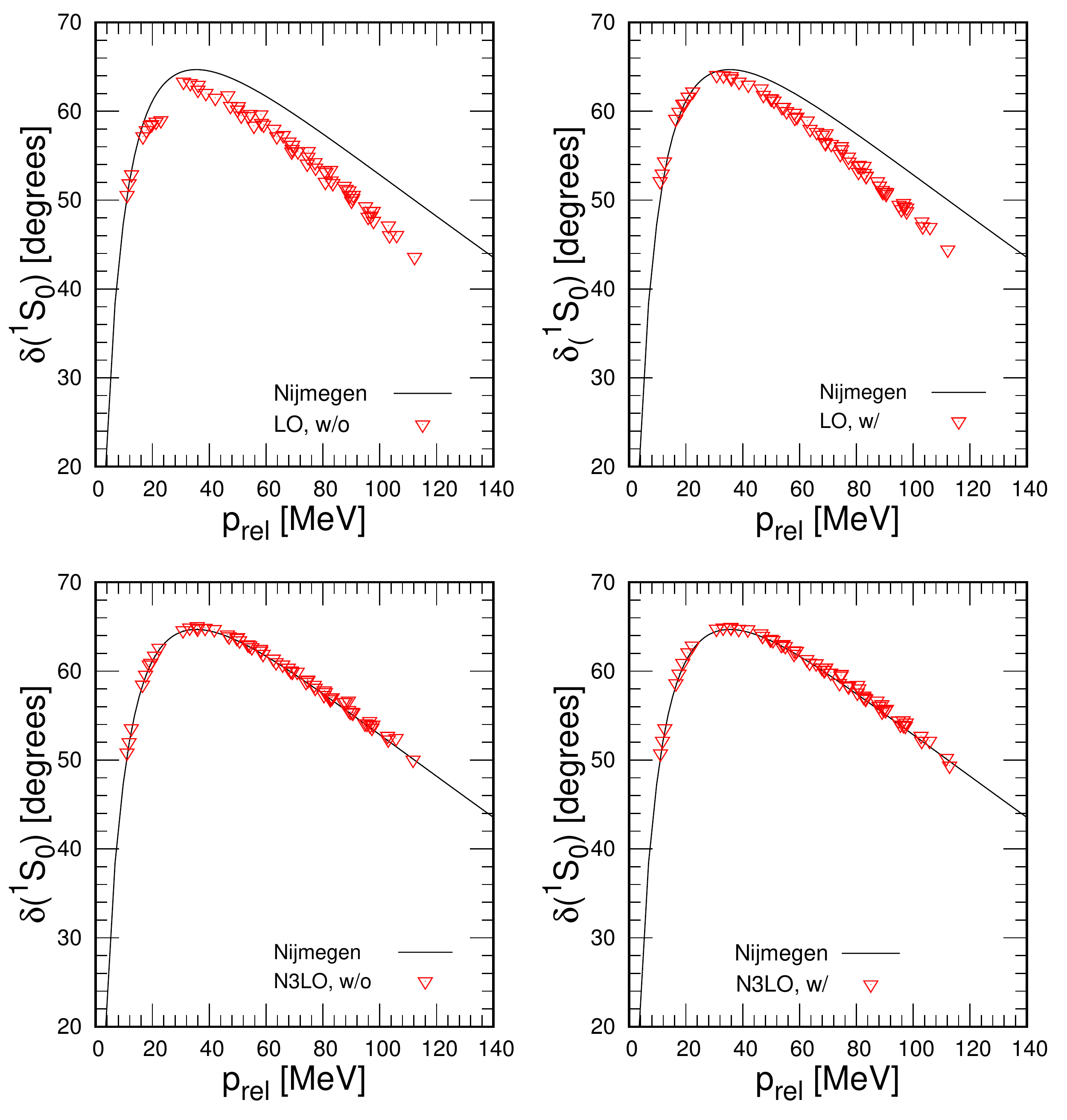}
\end{tabular}
\caption{(Color online) Neutron-proton scattering phase shifts of $^1S_0$ as a function of the relative momenta 
between the proton and neutron. The L{\"u}scher formula is used to extract the scattering phase shifts. 
 Top row: LO results, bottom row: N3LO results. `w/o' means without GIR corrections while `w/' denotes the results after restoring the 
 Galilean invariance. To study the finite volume effects, we did calculations using different size boxes, 
 $L = 14a, 16a$, $18a$ for the left two columns and $L = 24a, 26a$, $28a$ for the right two columns.
 In the generalization of the L{\"u}scher's formula to the non-rest frames, the symmetry of the subgroup of the cubic group is applied. 
 However, this symmetry is broken due to the breaking of the Galilean invariance. This leads to the rapid change of the phase shifts 
 at chiral LO. We can see that this behavior goes away after the Galilean invariance is restored. } 
\label{phaseShifts1S0}
\end{figure}

\begin{figure}
\centering
\begin{tabular}{cc}
\includegraphics[width=0.45\textwidth]{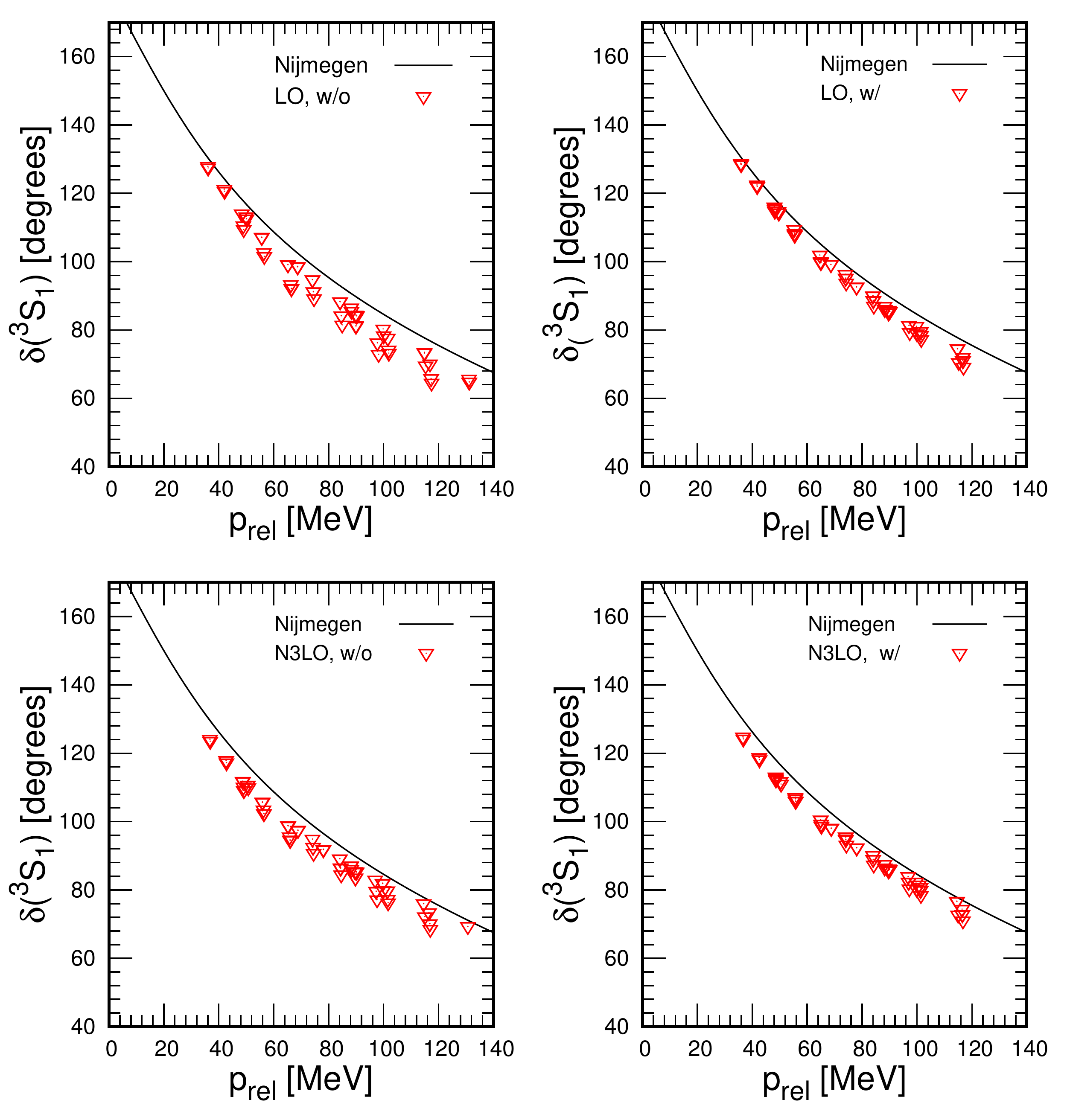} &
\includegraphics[width=0.45\textwidth]{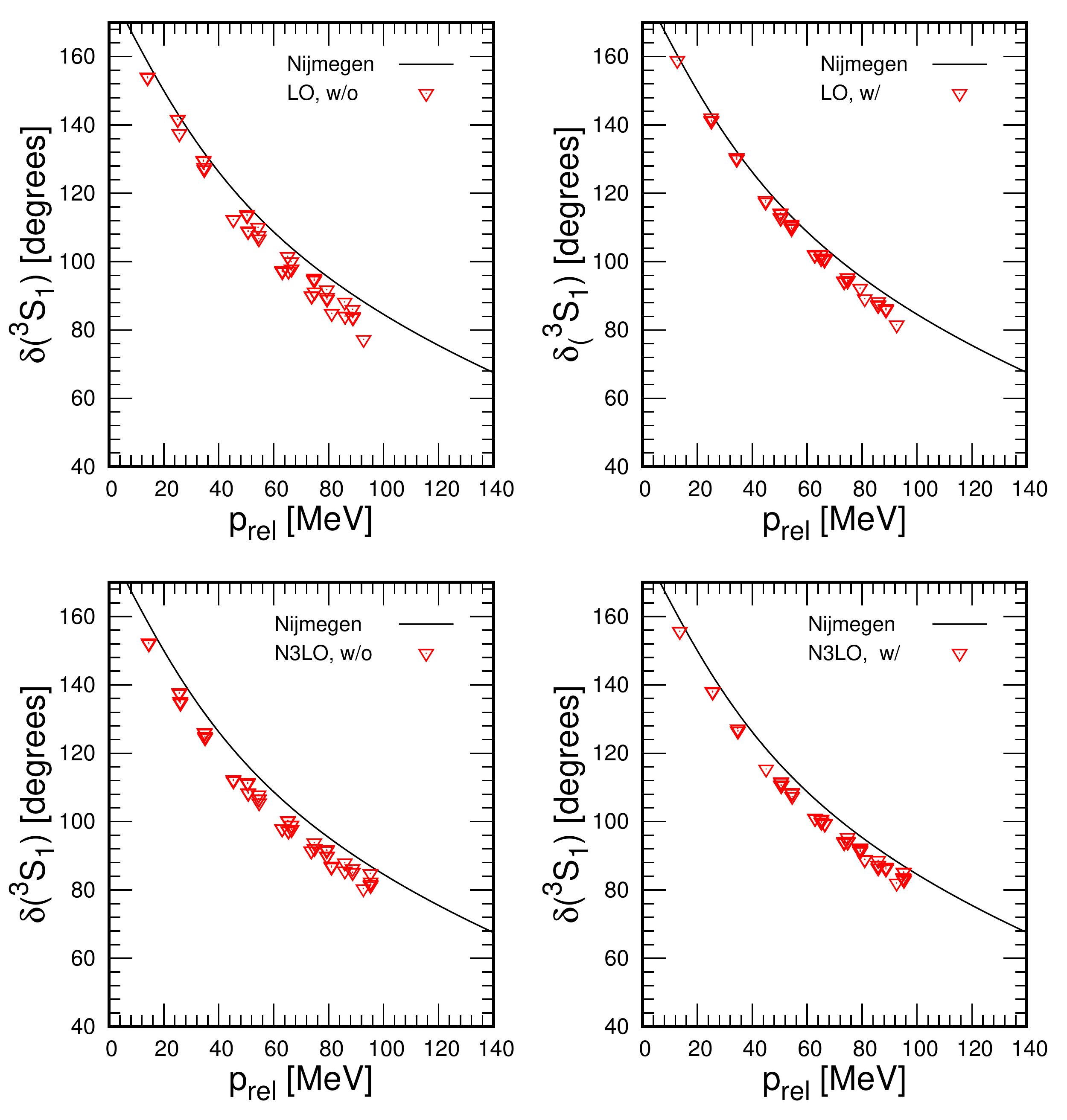} \\
\end{tabular}
\caption{(Color online) Neutron-proton scattering phase shifts of $^3S_1$ as a function of the relative momenta 
between the proton and neutron. The L{\"u}scher formula is used to extract the scattering phase shifts. 
`w/o' means without GIR corrections while `w/' denotes the results after restoring the 
 Galilean invariance. 
 To study the finite volume effects, we did calculations using different size boxes, 
 $L = 14a, 16a$, $18a$ for the left two columns and $L = 24a, 26a$, $28a$ for the right two columns.
 Top row: LO results, bottom row: N3LO results. } 
\label{phaseShifts3S1}
\end{figure}

We first calculate the neutron-proton scattering phase shifts for the $S$-wave channels, ${}^1S_0$ and ${}^3S_1$,  
using L{\"u}scher's formula. 
In order to obtain results for a wide energy range, we use several cubic boxes with volumes 
$V = (14a)^3$, $(16a)^3$,  and $(18a)^3$. To study the finite volume effects, larger cubic boxes with volume of $V = (24a)^3$, $(26a)^3$ and 
$(28a)^3$ are also used for the same calculations.
We first perform the calculation in the rest-frame, and then boost the proton-neutron system to moving frames with momenta 
${\bf P} = (2\pi/L) {\bf k}$. 
The results for ${}^1S_0$ and ${}^3S_1$ are shown in Figs.~\ref{phaseShifts1S0} and \ref{phaseShifts3S1}, respectively. 
The plots in top row are the LO results while those in the bottom row are the N3LO results. The left two columns are the results using the smaller boxes whereas 
the right two columns are the results using the larger boxes. `w/o' means 
without GIR corrections whereas `w/' denotes the results after restoring the Galilean invariance.  

From the plots in the top row of Fig.~\ref{phaseShifts1S0}, one can see that  there is clear GIB at LO although the calculation shows very good dispersion 
relation. The GIB of ${}^1S_0$ at LO  appears at low momenta, that is for relative momenta between 20 and 40~MeV. 
The Galilean invariance is restored after including the GIR corrections. It is necessary to mention that the deviation of the lattice results from those of the Nijmegen partial wave analysis is just because these are the LO results. At N3LO, it shows negligible GIB for ${}^1S_0$, which is consistent with
what the dispersion relation indicates. The case for ${}^3S_1$ is  different since the ground state of ${}^3S_1$ is a bound state. 
Both the LO and N3LO results show very small GIB. Combining the results of ${}^1S_0$ and ${}^3S_1$, we find that the N3LO interaction has 
less GIB than the LO interaction. This is because the GIB from the  non-locally smeared contact interactions at NLO and N3LO accidentally cancel some GIB effects 
caused by the lattice artifacts due to the nonzero lattice spacing. 

\begin{figure}
\centering
\includegraphics[width=0.6\textwidth]{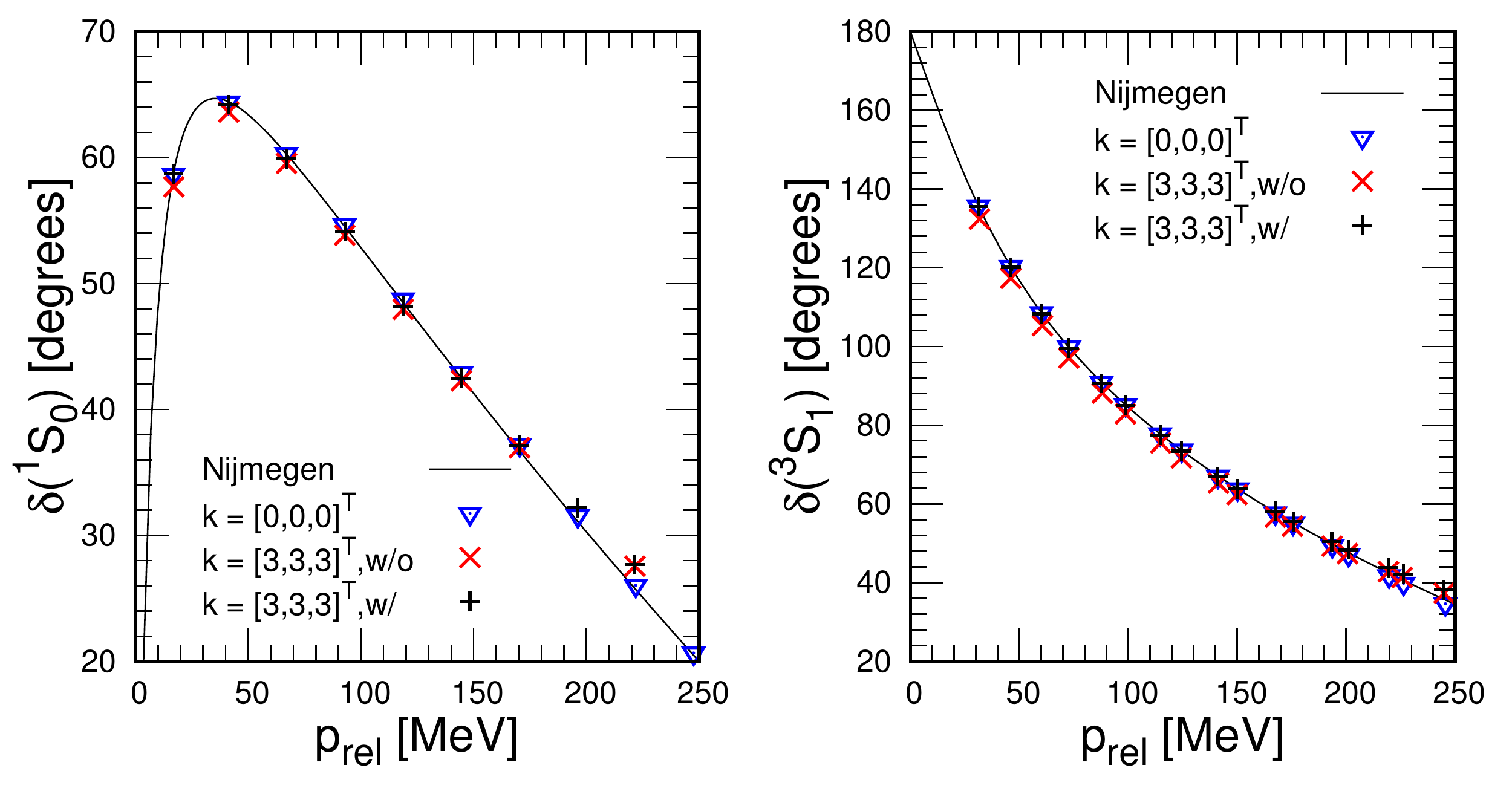}
\caption{(Color online)  $S$-wave neutron-proton scattering phase shifts  calculated using the spherical 
wall method as a function of the relative momenta between the two nucleons. }
\label{phaseShiftsS}
\end{figure}

We also calculate the scattering phase shifts for ${}^1S_0$ and ${}^3S_1$ using the spherical wall method.
The spherical wall method works with a one-dimension radial Hamiltonian matrix instead of a full three-dimension matrix. 
Thus the calculation is much faster than using L{\"u}scher's formula. Meanwhile, in order to reach the region where 
the NN interaction is vanishing a much larger box should be used. 
 In our calculation, we set $L = 40$ corresponding to radial distance
to be $La/2 = 26.4$~fm for $a = 1.32$~fm. To obtain a clear signal of GIB, we boost the proton-neutron system to 
moving frame with momentum ${\bf P} = (2\pi/L){\bf k}$ with ${\bf k} = [3, 3, 3]^T$. The N3LO results are shown in 
Fig.~(\ref{phaseShiftsS}). The small difference of the phase shifts in the two frames with ${\rm k} = [0,  0, 0]^T$ 
and ${\rm k} = [3, 3, 3]^T$ indicates the Galilean invariance breaking of the interaction is small.  
Additionally, one also observes small difference of the phase shifts for ${}^3S_1$ calculated using the spherical wall method 
from those calculated using the L{\"u}scher's formula. This is because there is a systematic error arising from 
the unphysical coupling of the $l = 0$ state with $l = 4$, $6$, and even higher partial waves using the generalized 
L{\"u}scher's formula in frames with ${\bf P} \ne {\bf 0}$~\cite{Doring:2012eu,Kim:2005gf,Briceno:2012yi}. 

\subsection{Mixing angles, $\epsilon_1$ and $\epsilon_2$,  and neutron-proton scattering phase shifts for $P$ and $D$ waves }

\begin{figure}[htp]
\centering
\includegraphics[width=0.6\textwidth]{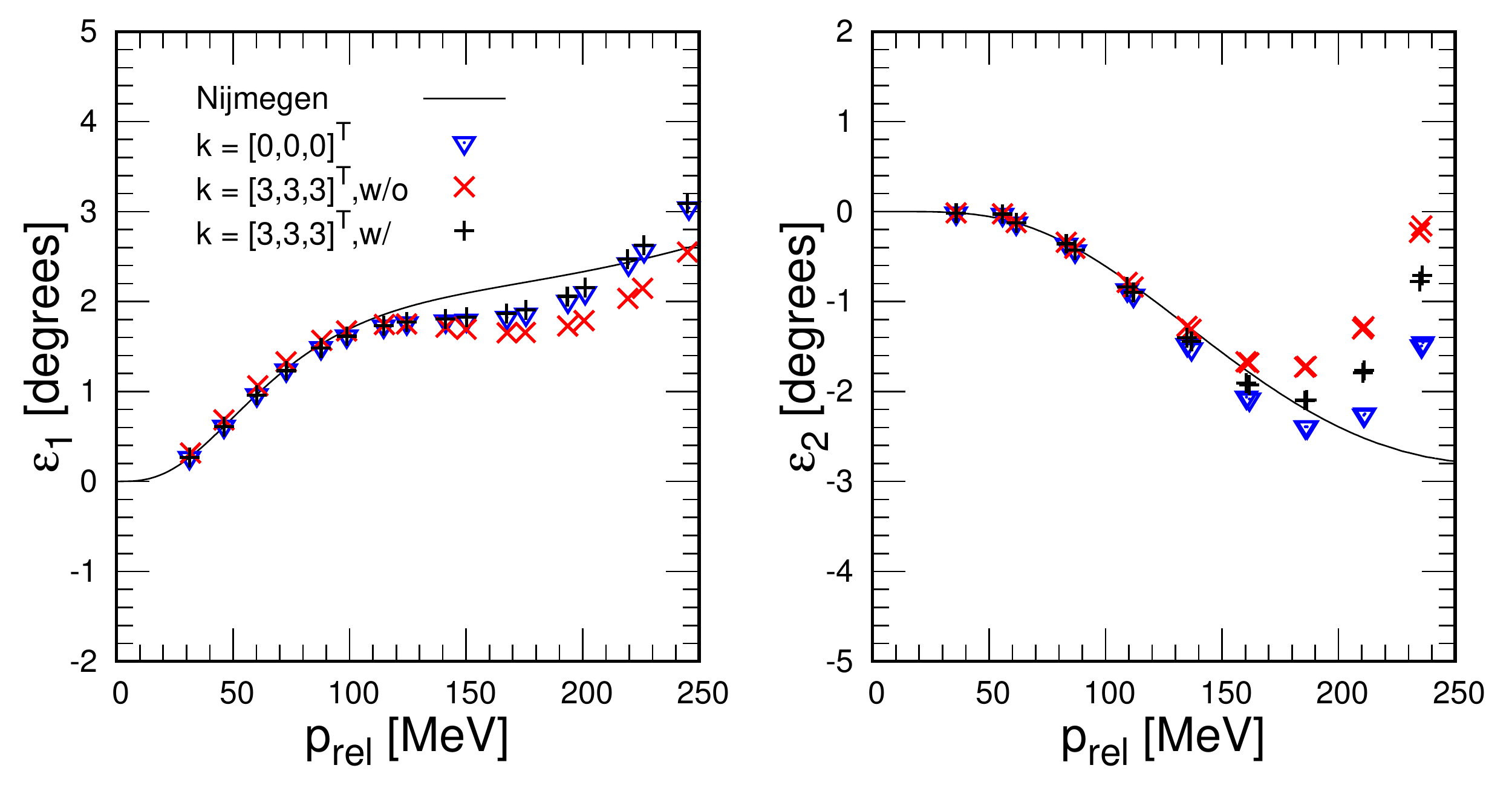}
\caption{(Color online) Mixing angles, $\epsilon_1({}^3S_1-{}^3D_1)$ and $\epsilon_2({}^3P_2-{}^3F_2)$, 
as a function of relative momenta between the proton and neutron. The spherical wall method is used. }
\label{mixing}
\end{figure}

\begin{figure}[t]
\centering
\includegraphics[width=0.6\textwidth]{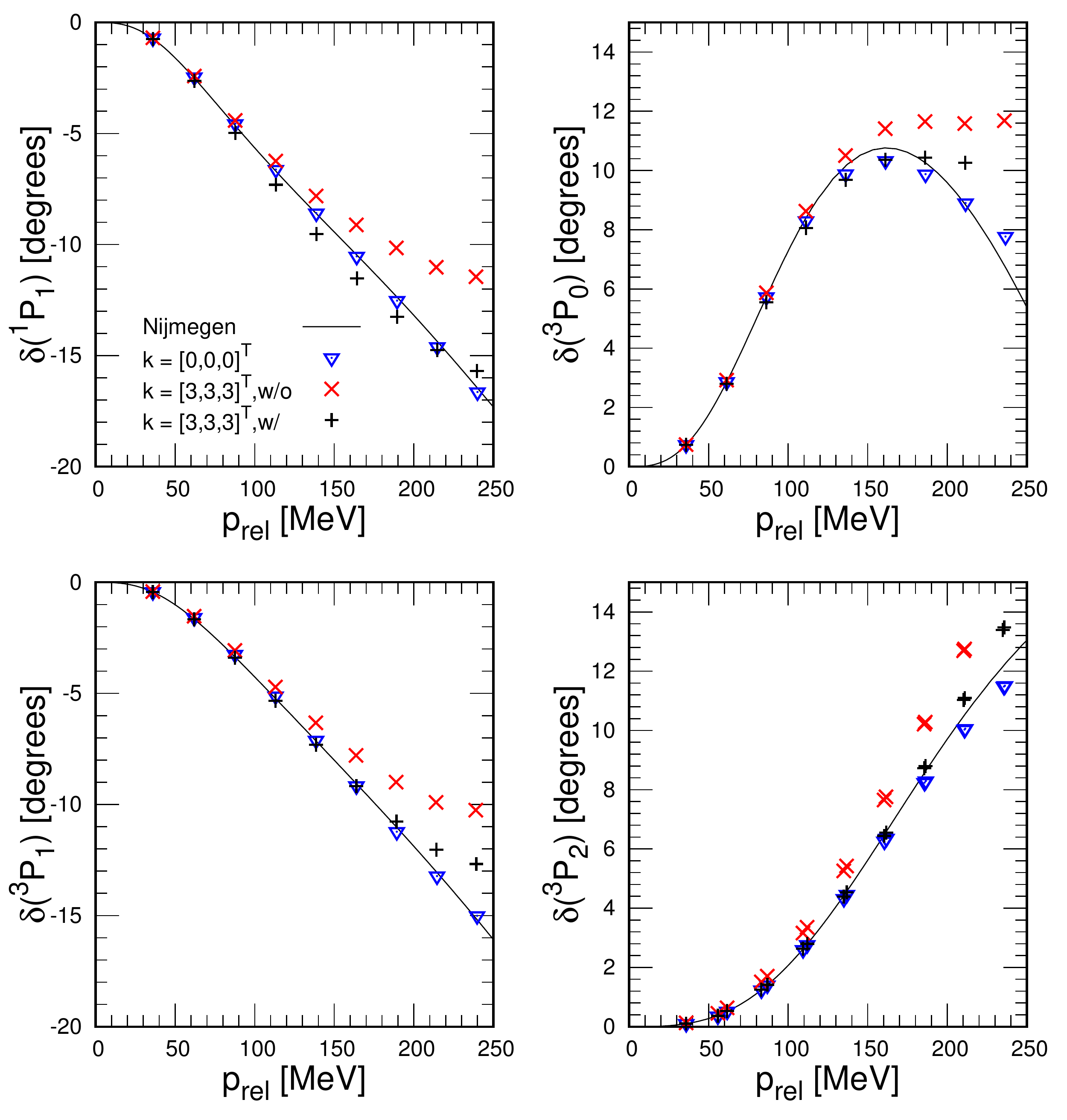}
\caption{(Color online) $P$-wave neutron-proton scattering phase shifts as a function of relative momenta 
between the proton and neutron. The spherical wall method is used. }
\label{phaseShiftsP}
\end{figure}

\begin{figure}
\centering
\includegraphics[width=0.6\textwidth]{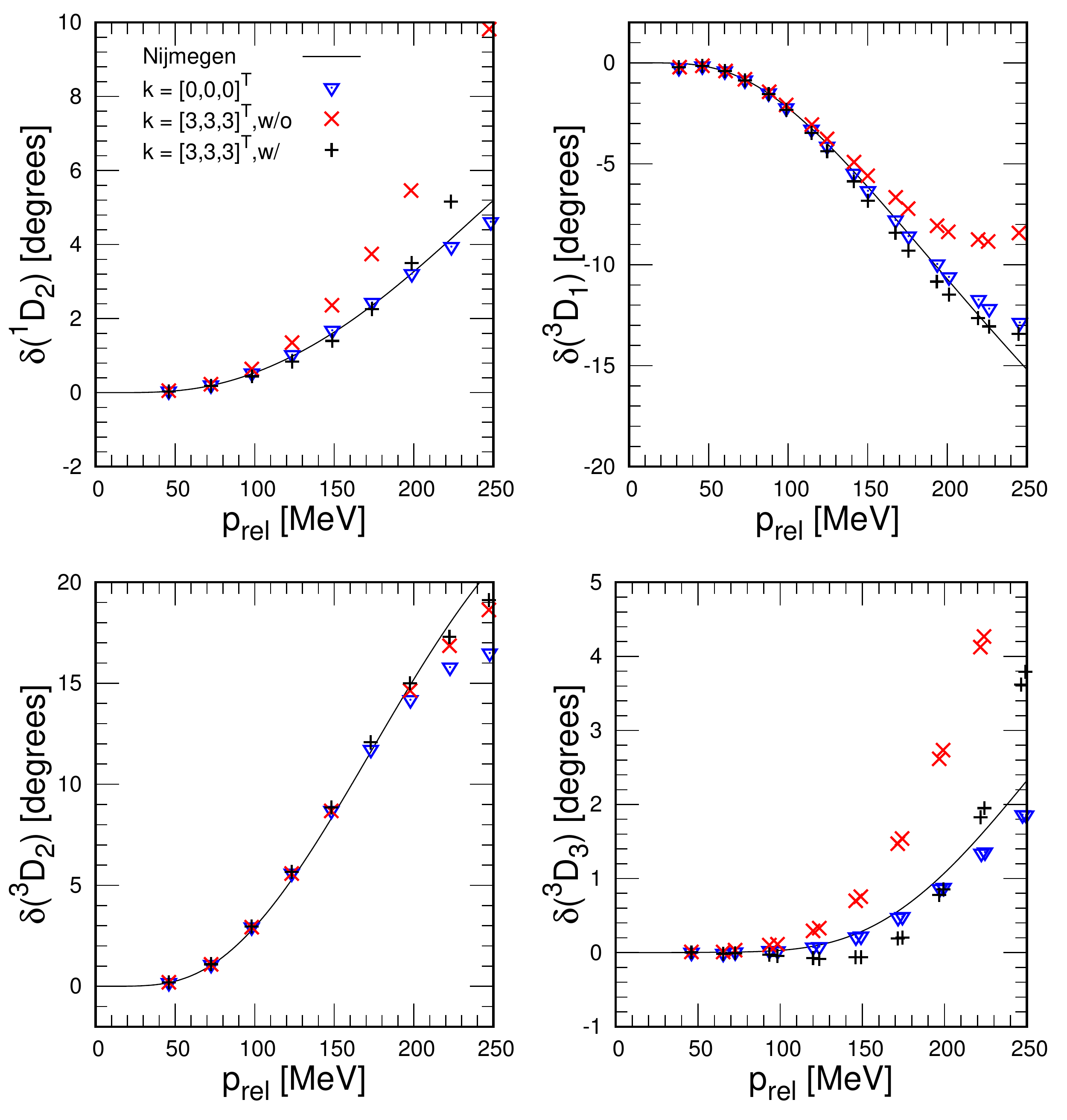} 
\caption{(Color online) $D$-wave neutron-proton scattering phase as a function of 
the relative momenta between the proton and neutron. The spherical wall method is used.}
\label{phaseShiftsD}
\end{figure}

As the  L{\"u}scher formula works well for the $S$ waves but not as accurately for the $P$, $D$ and even higher partial waves, 
we continue to calculate the mixing angles, $\epsilon_1({}^3S_1 - {}^3D_1)$ and $\epsilon_2({}^3P_2 - {}^3F_2)$, 
and proton-neutron scattering phase shifts for $P$ and $D$ waves  using the spherical wall method. 
The results are shown in Figs. ~\ref{mixing}, \ref{phaseShiftsP}, and \ref{phaseShiftsD}, respectively. 

From the plots, the Galilean invariance breaking for $\epsilon_1$ starts around $p_{\rm rel} = 120$~MeV while 
that for $\epsilon_2$ starts around $p_{\rm rel} = 150$~MeV. For $\epsilon_1$, after including the GIR correction 
the Galilean invariance is restored for the whole range $p_{r\rm rel} \leq 250$~MeV. For $\epsilon_2$, the GIR correction 
reduces the GIB very much although not completely.

The behavior of  the phase shifts for all four $P$-wave channels is very similar. The GIB appears in the high-momenta region,
 and starts around $p_{\rm rel} = 120$~MeV. After including the GIR correction, the GIB is largely removed. Very similarly,
 GIB also appears in high-momentum region for the $D$ waves. It starts around $p_{\rm rel} = 100$~MeV for 
${}^1D_2$ and ${}^3D_3$, and around  $p_{\rm rel} = 150$~MeV for ${}^3D_1$ and ${}^3D_2$. The GIR correction increases 
the starting points of GIB to around $p_{\rm rel} = 200$~MeV. 

\section{CONCLUSIONS}\label{summary}

With the rapid development of the high performance computers, nuclear lattice effective field theory has become a powerful tool 
in {\it ab initio} calculations of few- and many-body systems. However, getting efficient and precise nuclear forces on the lattice  
is more difficult than in the continuum due to the 
lattice artifacts caused by the nonzero lattice spacing. In order to reduce the lattice artifacts, in \cite{Elhatisari:2016owd} 
non-locally smeared contact operators were introduced. With only a few parameters, the binding energy of nuclei 
with nucleons up to twenty can be produced with good precision. However, the Galilean invariance is broken 
due to the nonlocal smearing parameter $s_{\rm NL}$ used to construct the contact operators. Another source 
of Galilean invariance breaking arises from the lattice itself. 

We investigate the effect of Galilean invariance breaking and restore the Galilean invariance on the lattice by 
studying the dispersion relation and proton-neutron scattering phase
shifts. We find that analyzing the phase shifts
in different frames is useful to detect GIB for the 
${}^1S_0$ partial wave while the dispersion relation
provides a more efficient tool in the 
${}^3S_1$ channel. This is because 
the ${}^1S_0$ ground state is a continuum state while the ground state of ${}^3S_1$ is a bound state. 

We find that the Galilean invariance breaking caused by the nonlocal smearing parameter $s_{\rm NL}$ partially cancels
that caused by the lattice artifacts due to the nonzero lattice spacing. Due to this cancellation, the Galilean invariance 
breaking of the NN interaction at N3LO is small. After including the GIR correction the Galilean invariance is restored. 

Our previous study shows that the non-locally smeared contact operators are promising in generating the 
binding of nucleons in nuclei.  The present study shows the Galilean invariance breaking is small, and the Galilean invariance 
can be restored after including the Galilean invariance restoration corrections. This interaction has been used in
Monte Carlo simulations for the nuclear binding of the light- and medium-mass nuclei.  We hope to be able to 
report the corresponding results in the new future.

\section*{Acknowledgements}

We acknowledge partial financial support from the Deutsche Forschungsgemeinschaft (SFB/TRR~110, ``Symmetries
and the Emergence of Structure in QCD''), the BMBF (Grant
No. 05P15PCFN1),
the U.S. Department of Energy (DE-SC0018638), and the Scientific and Technological Research 
Council of Turkey (TUBITAK project no. 116F400). Further support was provided by the Chinese 
Academy of Sciences (CAS) President's International Fellowship Initiative (PIFI) (grant no. 2018DM0034) 
and by VolkswagenStiftung (grant no. 93562).  The computational resources were provided by the J\"ulich 
Supercomputing Centre at Forschungszentrum J\"ulich, Oak Ridge Leadership Computing Facility, 
RWTH Aachen, North Carolina State University, and Michigan State University.

\begin{appendix}

\section{Lattice operator definitions}\label{LatticeDefinition}

The pertinent lattice operators were already defined in Ref.~\cite{Li:2018ymw}. However, for completeness, we list them again here.
With the dressed annihilation operator $a_{i, j}^{s_{\rm NL}}$, we define the pair annihilation operators $[a({\bf n})a({\bf n'})]^{s_{\rm
NL}}_{S,S_z,I,I_z}$, where
\begin{equation}
 [a({\bf n})a({\bf n'})]^{s_{\rm
NL}}_{S,S_z,I,I_z}=\sum_{i,j,i',j'} a^{s_{\rm NL}}_{i,j}({\bf n})M_{ii'}(S,S_z)M_{jj'}(I,I_z)a^{s_{\rm
NL}}_{i',j'}({\bf n'})
\label{spin-isospin}
\end{equation}
with
\begin{equation}
M_{ii'}(0,0)=\frac{1}{\sqrt{2}}[\delta_{i,0}\delta_{i',1}-\delta_{i,1}\delta_{i',0}],
\end{equation}
\begin{equation}
M_{ii'}(1,1)=\delta_{i,0}\delta_{i',0},
\end{equation}
\begin{equation}
M_{ii'}(1,0)=\frac{1}{\sqrt{2}}[\delta_{i,0}\delta_{i',1}+\delta_{i,1}\delta_{i',0}],
\end{equation}
\begin{equation}
M_{ii'}(1,-1)=\delta_{i,1}\delta_{i',1}.
\end{equation}
We define the lattice finite difference operation $\nabla_l$ on a general
lattice function $f({\bf n})$ as
\begin{equation}
\nabla_lf({\bf n})=\frac{1}{2}f({\bf n}+{\bf \hat{l}})-\frac{1}{2}f({\bf
n}-{\bf \hat{l}}),
\end{equation}
where ${\bf \hat{l}}$ is the spatial lattice unit vector in the $l$ direction. 
It is also convenient to define the lattice finite difference operation $\nabla_{1/2,l}$
defined on points halfway between lattice sites, 
\begin{equation}
\nabla_{1/2,l}f({\bf n})=f({\bf n}+\frac{1}{2}{\bf \hat{l}})-f({\bf
n}-\frac{1}{2}{\bf \hat{l}}).
\end{equation} 
This operation is used solely  to define the Laplace operator,
\begin{equation}
{\bf \nabla}^{2}_{1/2} = \sum_l{\bf \nabla}^{2}_{1/2,l}~.
\end{equation}
Further, we define the solid harmonics
\begin{equation}
R_{L,L_z}({\bf r}) = \sqrt{\frac{4\pi}{2L+1}}r^L Y_{L,L_z}(\theta,\phi),
\end{equation}
and their complex conjugates
\begin{equation}
R^*_{L,L_z}({\bf r}) = \sqrt{\frac{4\pi}{2L+1}}r^L Y^*_{L,L_z}(\theta,\phi).
\end{equation}
Using the pair annihilation operators, lattice finite differences,
and the solid harmonics, we define
the operator
\begin{equation}
P^{2M,s_{\rm NL}}_{S,S_z,L,L_z,I,I_z}({\bf n}) =[a({\bf n}){\bf \nabla}^{2M}_{1/2}R^*_{
L,L_{z}}({\bf \nabla})a({\bf n})]^{s_{\rm
NL}}_{S,S_z,I,I_z}, 
\end{equation}
where ${\bf \nabla}_{1/2}^{2M}$ and ${\bf \nabla}$ act on the second
annihilation operator.  More explicitly stated, this means that we act on
the ${\bf n'}$ in Eq.~(\ref{spin-isospin}) and then set ${\bf n'}$ to equal
${\bf n}$.
The even integer $2M$ gives us higher powers of the finite differences. Writing the Clebsch-Gordan coefficients as $\langle S S_z L L_z \vert J J_z\rangle$,
we define 
\begin{equation}
O^{2M,s_{NL}}_{S,L,J,J_z,I,I_z}({\bf n})= 
\sum_{S_z,L_z}\langle S S_z L L_z \vert J J_z\rangle P^{2M,s_{NL}}_{S,S_z,L,L_z,I,I_z}({\bf
n}).
\end{equation}

\end{appendix}

\bibliography{GIR}{}
\bibliographystyle{apsrev4-1}

\end{document}